\documentclass[11pt]{article}
\usepackage{graphicx,natbib}
\def\lesssim{\hbox{\rlap{\hbox{\lower4pt\hbox{$\sim$}}}\hbox{$ <$}}}
\def\gtrsim{\hbox{\rlap{\hbox{\lower4pt\hbox{$\sim$}}}\hbox{$> $}}}

\begin{document}
\sloppypar

\thispagestyle{empty}

\begin{centering}
\parindent 0mm

{\Large\bf Constraints on the late X-ray emission from the
  low-energy~GRB 031203: INTEGRAL data} 

\vspace{0.5cm}
{\large\bf S.Y. Sazonov$^{1,2}$, A.A. Lutovinov$^{1,2}$,
  E.M. Churazov$^{1,2}$, and R.A. Sunyaev$^{1,2}$}
 
\vspace{0.5 cm}
$^{1}$ Space Research Institute, Russian Academy of Sciences,
Profsoyuznaya 84/32, Moscow 117997, Russia 

$^{2}$ Max-Planck-Institut f\"ur Astrophysik,
           Karl-Schwarzschild-Str. 1, D-85740 Garching, Germany

\end{centering}

\vspace{1cm}

\noindent{\large\bf Abstract} -- Comparison of the INTEGRAL upper
limits on the hard 
X-ray flux before and after the low-energy GRB~031203 with the XMM
measurements of the dust-scattered radiation at lower energies
suggests that a significant fraction of the total burst 
energy could be released in the form of soft X-rays at an
early afterglow stage with a characteristic duration of
$\sim$100­-1000~s. The overall time evolution of the GRB~031203 afterglow
may have not differed qualitatively from the behavior of standard
(i.e., more intense) bursts studied by the SWIFT observatory. The
 available data also admit the possibility that the dust-scattered
 radiation was associated with an additional soft component in the
 spectrum of the gamma-ray burst itself. 

\clearpage

\section{Introduction}

GRB~031203, discovered on December 3, 2003, by the INTEGRAL
observatory (Mereghetti and G\"otz 2003), has attracted much attention
because the total energy released during this burst (assuming the
radiation to be isotropic) was approximately three orders of magnitude
lower than that for standard GRBs. At the same time, the time profile
and energy spectrum of the burst were quite normal (Sazonov et
al. 2004). The afterglow of this GRB was also weak (Soderberg et
al. 2004). GRB~031203 is similar in these properties to another burst, 
GRB~980425, which was believed to be a unique event before. The
similarity of these bursts is also confirmed by the fact that each of
them was identified with the explosion of a supernova associated with
the core collapse of a high-mass star (Malesani et al. 2004; Galama et
al. 1998). Since both bursts were discovered only owing to their
relative proximity (GRB~980425 at redshift $z=0.0085$, Tinney et
al. 1998; GRB 031203 at $z=0.106$, Prochaska et al. 2004), it was
suggested that such low-energy bursts could be a more common cosmic
phenomenon than standard high-energy bursts (Sazonov et al. 2004;
Prochaska et al. 2004). 

Additional interest in GRB~031203 is related
to the detection of an expanding X-ray halo around the burst position
in the sky by the XMM-Newton observatory on the first day after the
GRB. This halo was in the shape of two concentric rings. A natural
explanation is that this is the soft X-ray radiation from GRB~031203
arrived with a considerable time delay after its scattering by two
layers of Galactic dust at a distance of $\sim$1~kpc from the Earth (Vaughan
et al. 2004). A careful analysis of the data on the evolution of the
halo showed that the X-ray blast that produced it occurred no earlier
than $\sim$100~s before and no later than $\sim$1300~s after the GRB (Watson et
al. 2006). The estimate (Watson et al. 2006; Tiengo and Mereghetti
2006) of the soft X-ray fluence (at energies $\sim$1~keV) was several times
higher than the value obtained by extrapolating the spectrum of the
GRB itself that was measured by the INTEGRAL observatory at energies
above 17~keV (Sazonov et al. 2004). This raises the question of the
origin of the additional soft X-ray radiation. In this paper, we
attempt to approach the solution of this question using the upper
limits on the X-ray flux before and after GRB 031203~derived from
INTEGRAL observations. 

\section{Observations}

We analyzed the data from the IBIS/ISGRI detector (Ubertini et
al. 2003) aboard the INTEGRAL observatory (Winkler et al. 2003) that 
covered the from 21 h 32 min December 3 to 22 h 01 min December 4,
2003, (UT), i.e., half an hour before and a day after the burst. Most
of the data were obtained during observations with a stable
orientation ($\sim$30~min in duration), which alternated with two-minute
periods of spacecraft slewing (within a few degrees, so the GRB
position always remained within the field of view). Since the
IBIS/ISGRI data recording was frequently interrupted during slewing,
we also analyzed the data from the SPI spectrometer (with a lower
sensitivity than IBIS/ISGRI) for these intervals. This ensured a
continuous monitoring of the position of GRB~031203 throughout the
period under consideration. 

To reconstruct the X-ray light curve of
GRB~031203 from the IBIS/ISGRI data, we constructed images of the sky
for time intervals from several seconds to one day. We also searched
for fast variability (on time scales $\sim$1~s) of the detector count rate
within several minutes before and after the burst. Both types of
analysis revealed no X-ray radiation from GRB~031203 in the energy
ranges 17­-25~keV (i.e., just above the detector sensitivity threshold)
and 17­-60~keV, except for the $\sim$40-s-long burst itself (Figs.~1 and
2). The data on the burst itself were published previously (Sazonov et
al. 2004). Given that the observing conditions almost always remained
invariable and assuming the radiation spectrum to be identical to that
of the Crab Nebula, we can place (3$\sigma$) upper limits on the mean flux
from GRB~031203 for interval $\Delta t$ [s], excluding the GRB and the
observatory slewing periods: $\sim 6\times 10^{-9}(\Delta
t)^{-1/2}$~erg~cm$^{-2}$~s$^{-1}$ (17­-25~keV) and $\sim 8\times
10^{-9} (\Delta t)^{-1/2}$~erg~cm$^{-2}$~s$^{-1}$ (25­-60~keV). The
corresponding SPI flux limits during slewing are approximately a
factor of 5 worse.

The long-term light curve shown in Fig.~2 covers the period in which two
series of XMM observations were performed, beginning 6~h after the
GRB. Apart from the expanding X-ray halo, these observations revealed
a decaying ($f_{\rm X}\propto t^{-\alpha}$, $\alpha=0.55\pm 0.05$) X-ray
afterglow of the GRB (Watson et al. 2004). We see from the figure that
the INTEGRAL upper limits on the hard X-ray flux are consistent with
the XMM flux from the X-ray afterglow at its late stage ($t > 6$~h). 

The INTEGRAL upper limits on the hard X-ray flux for the last $\sim$100~s
before the GRB and the first $\sim$1300~s after it (Fig. 1) are of greatest
interest, since the bulk of the soft X-ray fluence, which was
estimated by Watson et al. (2006) to be $F_{\rm X}=(2.1\pm0.4)\times
10^{-6}$~erg~cm$^{-2}$~keV$^{-1}$ at energy 1~keV, was released
precisely in this period. The radiation spectrum in the energy range
0.7­-6~keV was much softer (a power law with a photon index
$\Gamma_{\rm X}=2.0\pm 0.15$) than the spectrum of the GRB itself at energies
above 17~keV ($\Gamma=1.63\pm 0.06$, see Sazonov et al. 2004). However, it
should be noted that the estimated soft X-ray fluence from GRB~031203
is inversely proportional to the optical absorption $A_V$ in the dust
layers with which the echo is associated. The above estimate was
obtained for $A_V =2.0$, which is considered to be the most probable
one. If the maximum admissible value of $A_V\sim 2.6$ is taken, then the
fluence will decrease by $\sim$30\% (Watson et al. 2006). An even greater
uncertainty is related to the size distribution and other parameters
of the dust grains, i.e., to the coefficient $\tau/A_V$ at energies
$\sim$1~keV, where $\tau$ is the optical depth of the scattering dust
(see, e.g., Draine 2003). This is why Tiengo and Mereghetti (2006)
obtained a soft X-ray fluence from GRB~031203 that is approximately a
factor of 4 lower than that obtained by Watson et al. (2006) for the
same $A_V =2$ by independently analyzing the same XMM data: $F_{\rm
  X}=(3.6\pm 0.2)\times 10^{-7}$~erg~cm$^{-2}$ in the energy range
1­-2~keV. Almost the same value was obtained for the spectral slope:
$\Gamma_{\rm X}=2.1\pm 0.2$.  

Taking into account the significant uncertainty related to the
coefficient $\tau/A_V$, we may consider the fluences $F_{\rm X}$ from
the above two papers as the upper and lower limits on the actual
value. These limits are shown in Fig.~3 in the same plot together with
the INTEGRAL data; more specifically, the GRB spectrum at energies
above 17~keV and the upper limits on the fluence in the energy ranges
17­-25 and 25­-60~keV for any episode with duration $\Delta t=10$,
100, and 1000~s before or after the burst. It follows from Fig.~3 that
if the soft X-ray radiation was released during the GRB, then the GRB
spectrum must exhibit an additional soft component (and a deep minimum
in the energy range from $\sim$6 to $\sim$17~keV if the flux
normalization of Watson et al. 2006 is correct).

We may also assume that the soft X-ray radiation was released not
during the main GRB phase. In this case, we must take into account the
fact that the soft X-ray fluence estimated from the observations of
the dust echo naturally includes the fraction related to the X-ray
radiation during the GRB. This fraction can be estimated by extending
the power-law spectrum measured by the IBIS/ISGRI detector to the low
energies. Subtracting this fraction changes the estimate of Watson et
al. (2006) only slightly: the fluence near 1~keV decreases by $\sim$10\%,
while the effective spectral slope of the soft X-ray radiation
increases to $\Gamma_{\rm X}\sim 2.1$. The extrapolation of this
spectrum to the high energies (Fig.~3) passes well above the
IBIS/ISGRI upper limits on the fluence in the energy ranges 17-­25 and
25-­60~keV for $\Delta t\sim 1000$~s and shorter intervals. At the
same time, subtracting the GRB contribution decreases the fluence
obtained by Tiengo and Mereghetti (2006) by $\sim$30\% and increases
the effective spectral slope of the soft X-ray radiation to
$\Gamma_{\rm X}\sim 2.4$. This single-power-law spectrum agrees well
with our upper limits even at $\Delta t\gtrsim 100$~s. 

We cannot rule out the possibility that an
X-ray flare with a duration of $\lesssim 100$~s occurred in the period from 300
to 416~s after the burst, when the INTEGRAL observatory was
slewed. There are relatively weak SPI limits on the hard X-ray fluence
for this interval. As in the previous case, the soft X-ray fluence and
the spectrum obtained by Watson et al. (2006) can be reconciled with
these upper limits only if there is a knee in the spectrum in the
energy range 6­-17~keV. At the same time, the estimate obtained by
Tiengo and Mereghetti (2006) does not come into conflict with the SPI
limits if the single-power-law spectrum with a slope of $\Gamma_{\rm
  X}\sim 2.4$ is extended above 17~keV (see Fig.~3). 

\section{Discussion}

As we noted above, it may well be that the soft X-ray radiation from
which the echo on dust was observed in our Galaxy was generated during
GRB 031203 itself. This would imply that the GRB spectrum contained an
additional soft component at energies below 17~keV. This is atypical
of standard GRBs (with a much higher total energy than GRB~031203),
whose simultaneous soft and hard X-ray observations usually reveal a
softening of the spectrum above a certain energy (Strohmayer et
al. 1998; Frontera et al. 2000), as well as for another well-known
low-energy burst, GRB~980425, whose spectrum was directly measured in
the energy range 2­-700~keV by the BeppoSAX observatory (Frontera et
al. 2000). 

The INTEGRAL and XMM data are consistent with the fact that
the bulk of the soft X-ray fluence was released during a prolonged
event ($\Delta t\sim$100­-1000~s) that occurred slightly later than
the GRB. This hypothesis encounters no difficulties if the soft X-ray
fluence was close to the value obtained by Tiengo and Mereghetti
(2006). In this case, the spectrum could be a single power law with a
slope $\Gamma_{\rm X}\sim$2­-2.5 up to high energies (above
17~keV). If, however, the soft X-ray fluence was closer to the value
obtained by Watson et al. (2006), then the 6­-17~keV spectrum must
steepen sharply toward the high energies; otherwise the derived upper
limits on the hard X-ray flux outside the burst would be exceeded. 

Such a prolonged episode ($\Delta t\gtrsim 100$~s) could be
associated both with the initial afterglow stage and with an
additional, X-ray pulse that followed the GRB. In the last year, the
SWIFT observatory has made it possible to systematically study the
soft X-ray radiation from GRB sources beginning from $\sim 100$~s after their
detection. In several cases, intense prolonged X-ray flares were
actually observed a few minutes after the GRB; these may have been
associated with ongoing activity of the central engine (Burrows et
al. 2005). Such a scenario is not ruled out for GRB~031203 either, as
was noted by Watson et al. (2006). However, most of the GRBs observed
by SWIFT have smooth X-ray light curves after $t\sim$100~s (where the time
is measured from the burst onset) with a fast flux decay ($f_{\rm
  X}\propto t^{-\alpha}$, $\alpha\sim 3$) in the first several minutes
followed by an almost flat segment ($\alpha\sim$0.2-­0.8), which again
gives way to a relatively fast decay several hours later
($\alpha\sim$1-­1.5) (Nousek et al. 2005). In many cases, at the 
initial decay stage, the radiation spectrum below 10~keV is much
softer than the spectrum of the GRB itself (measured above $\sim$15~keV by
the BAT instrument onboard the SWIFT observatory) and is generally
almost constant at the subsequent stages (Goad et al. 2006; Nousek et
al. 2005). The soft X-ray fluence released in the interval from $\sim100$~s
to several hundred seconds after the burst may account for an
appreciable fraction of the hard X-ray and gammaray fluence during the
burst (Chincarini et al. 2005).

It is possible that in the case of GRB~031203, there was a similar early
 ($t\lesssim 1000$~s) X-ray afterglow stage, when $\sim20$\% (if the
 estimate of Tiengo and Mereghetti (2006) is used) or slightly more of
 the total GRB was  released. This hypothesis is supported by the fact
 that a plateau ($\alpha\sim 0.5$, Watson et al. 2004) followed by a
 decay ($\alpha\sim 1.0$, Soderberg et al. 2004) similar to most of
 the bursts studied by the SWIFT observatory can be clearly
 distinguished in the light curve for the late ($t>6$~h) X-ray
 afterglow of GRB~031203. This scenario is also supported by the fact
 that the late X-ray afterglow of GRB~031203 had 
 approximately the same spectrum ($\Gamma_{\rm X}=1.90\pm 0.05$,
 Watson et al. 2004) as the early X-ray radiation from which the echo
 was observed.

It should be noted that a relatively slow X-ray flux decay ($\alpha< 1$,
Burenin et al. 1999; Tkachenko et al. 2000; Chincarini et al. 2005)
was detected in some of the bursts at an early ($t\lesssim 1000$~s) afterglow
stage. GRB~031203 may have also had a similar afterglow phase during
which the bulk of the soft X-ray fluence was released. However, this
phase should anyway be followed by a stage of fast flux decay, since,
according to the XMM data, the afterglow intensity 6~h after the burst
already fell by several orders of magnitude (see Fig.~2). 

\vspace{1cm}
\noindent{\bf Acknowledgments} We are grateful to S.V. Molkov and
M.G. Revnivtsev for helpful discussions. This work was supported by
the Russian Foundation for Basic Research. A.A.L. acknowledges the
financial support from the Russian Foundation for Support of Science.


\begin{figure}
\includegraphics[width=\columnwidth]{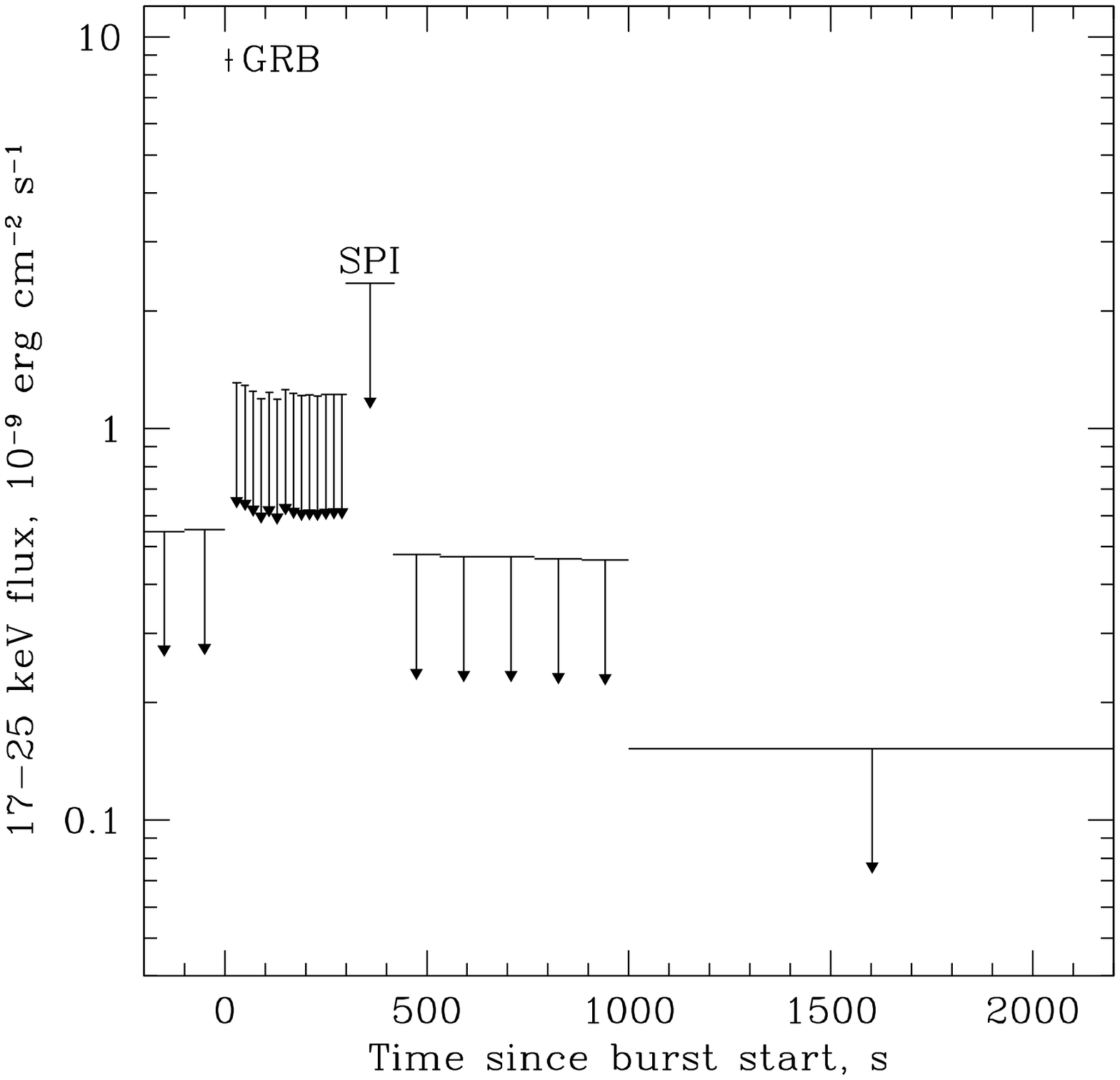} 
\caption{IBIS/ISGRI light curve of GRB 031203 in the energy range
  17­-25~keV in the period from -200 to 2200~s after the GRB onset. The
  upper limits imply that no flux was detected at a confidence level
  above 3$\sigma$. For the interval 300­-416~s in which the observatory was
  slewed, the SPI upper limit is shown. The point in the interval
  0­-20~s (labeled GRB) corresponds to the main GRB phase.
} 
\label{lc_short}
\end{figure}

\begin{figure}
\includegraphics[width=\columnwidth]{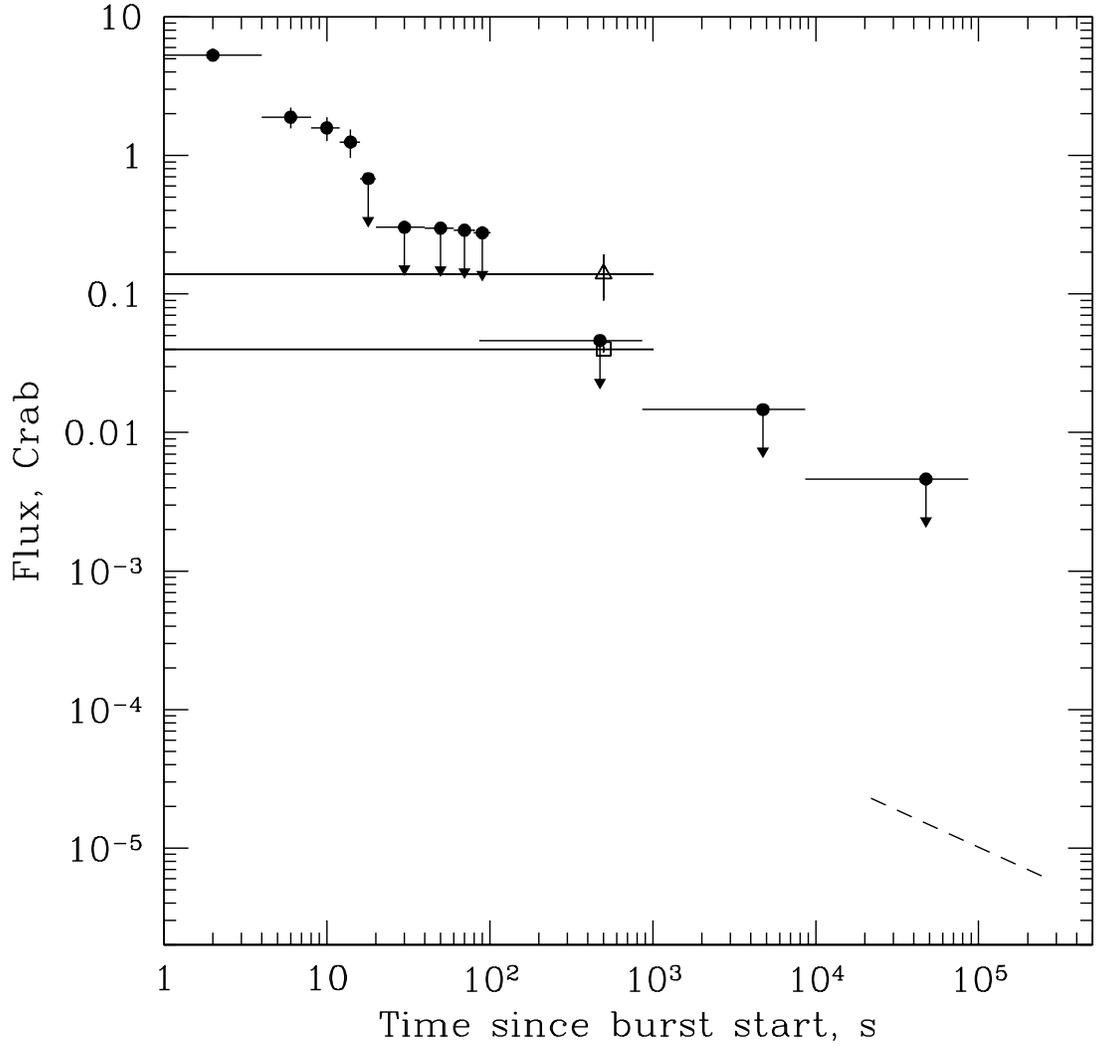} 
\caption{Light curve of GRB~031203 in the energy ranges 17­-25~keV
  (INTEGRAL data, circles) and $\sim$1­-5~keV (XMM data, triangle and square)
  in units of the flux from the Crab Nebula. Two different estimates
  of the soft X-ray fluence are shown: those from Watson et al. (2006)
  (triangle) and Tiengo and Mereghetti (2006) (square). These values
  were obtained by assuming that the X-ray event lasted for $\Delta t =1000$~s
  after the burst onset. The dashed line indicates a power-law decay
  of the X-ray afterglow ($f_{\rm X}\propto  t^{-0.55}$) at its late
  stage (XMM data, Watson et al. 2004). The measurements were corrected for
  interstellar absorption ($N_{\rm H}\sim 9\times 10^{21}$~cm$^{-2}$). 
}
\label{lc_long}
\end{figure}
 
\begin{figure}
\includegraphics[width=\columnwidth]{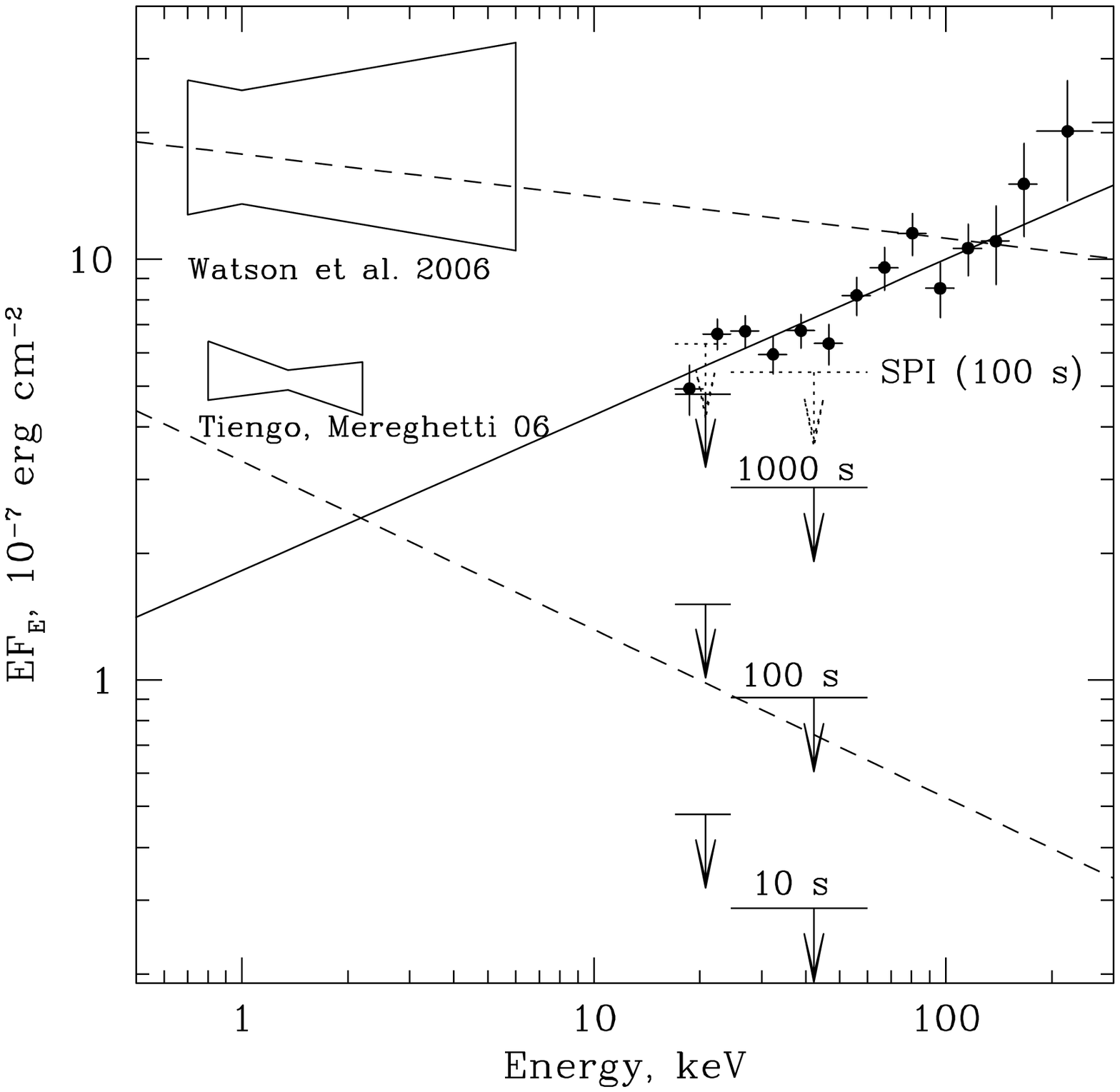} 
\caption{Constraints on the X-ray spectrum from GRB~031203, as derived
  from INTEGRAL and XMM data. The circles with error bars indicate the
  IBIS/ISGRI spectrum of the GRB (Sazonov et al. 2004). The solid line
  indicates the power-law ($\Gamma=1.63$) best fit to this spectrum extended
  to the low energies. Also shown are the IBIS/ISGRI (3$\sigma$) upper
  limits on the fluence in the energy ranges 17­-25 and 25-­60~keV for a
  time interval of 10, 100, and 1000~s outside the burst, the SPI
  limit for $\Delta t=100$~s, and the soft X-ray fluences estimated by Watson
  et al. (2006) and Tiengo and Mereghetti (2006). The dashed lines
  indicate the extrapolation of these spectra after the subtraction of
  the GRB-related radiation.
}
\label{spec_limits}
\end{figure}


\newpage
\begin{thebibliography}{}

\bibitem[\protect\citeauthoryear{Burenin et al.}{1999}]{bvg+99}
  R.A. Burenin, A.A. Vikhlinin, M.R. Gilfanov, et al.,
  Astron. Astrophys., 344, L53 (1999).

\bibitem[\protect\citeauthoryear{Burrows et al.}{2005}]{brf+05}
  D.N. Burrows, P. Romano, A. Falcone, et al., Science, 309, 1833
  (2005). 

\bibitem[\protect\citeauthoryear{Chincarini et al.}{2005}]{cmr+05}
G. Chincarini, A. Moretti, P. Romano, et al., astro-ph/0506453 (2005). 

\bibitem[\protect\citeauthoryear{Draine}{2003}]{d03} B.T. Draine,
  Astrophys. J., 598, 1026 (2003). 

\bibitem[\protect\citeauthoryear{Frontera et al.}{2000}]{fac+00}
F. Frontera, L. Amati, E. Costa, et al., Astrophys. J. Suppl., 127, 59 (2000). 

\bibitem[\protect\citeauthoryear{Galama et al.}{1998}]{gvv+98}
  T.J. Galama, P.M. Vreeswijk, J. van Paradijs, et al. Nature, 395, 670 (1998).

\bibitem[\protect\citeauthoryear{Goad et al.}{2005}]{gtp+05}
  M.R. Goad, G. Tagliaferri, K.L. Page, et al.,
  Astron. Astrophys. (2006) (in press).

\bibitem[\protect\citeauthoryear{Malesani et al.}{2004}]{mtc+04}
D. Malesani, G. Tagliaferri, G. Chincarini, et al., Astrophys. J.,
609, L5 (2004).

\bibitem[\protect\citeauthoryear{Mereghetti \& G\"otz}{2003}]{mg03}
S. Mereghetti, D. G\"otz, GCN Circ., 2460 (2003)

\bibitem[\protect\citeauthoryear{Nousek et al.}{2005}]{nkg+05}
  J.A. Nousek, D.C. Morris, D.N. Burrows, et al., astro-ph/0508332 (2005). 

\bibitem[\protect\citeauthoryear{Prochaska et al.}{2004}]{pbc+04}
J.X. Prochaska, J.S. Bloom, H.-W. Chen, et al., Astrophys. J., 611,
200 (2004).  

\bibitem[\protect\citeauthoryear{Sazonov et al.}{2004}]{sls04}
  S.Y. Sazonov, A.A. Lutovinov, R.A. Sunyaev, Nature, 430, 646 (2004).

\bibitem[\protect\citeauthoryear{Soderberg et al.}{2004}]{skb+04}
  A.M. Soderberg, S.R. Kulkarni, E. Berger, et al., Nature, 430, 648 (2004).

\bibitem[\protect\citeauthoryear{Strohmayer et al.}{1998}]{sfm+98}
  T.E. Strohmayer, E.E. Fenimore, T. Murakami, \& A. Yoshida,
  Astrophys. J., 500, 873 (1998). 

\bibitem[\protect\citeauthoryear{Tiengo \& Mereghetti}{2006}]{tm06}
  A. Tiengo \& S. Mereghetti, Astron. Astrophys. 2006 (in press). 

\bibitem[\protect\citeauthoryear{Tinney et al.}{1998}]{tsc+98}
  C. Tinney, R. Stathakis, R. Cannon, et al., IAU Circ. 6896 (1998). 

\bibitem[\protect\citeauthoryear{Tkachenko et al.}{2000}]{tts+00}
A.Y. Tkachenko, O.V. Terekhov, R.A. Sunyaev, et al.,
Astron. Astrophys., 358, L41 (2000).

\bibitem[\protect\citeauthoryear{Ubertini et al.}{2003}]{uld+03}
P. Ubertini, F. Lebrun, G. Di Cocco, et al., Astron. Astrophys., 411,
L131 (2003).

\bibitem[\protect\citeauthoryear{Vaughan et al.}{2004}]{vwo+04}
  S. Vaughan, R. Willingale, P.T. O'Brien, et al., Astrophys. J., 603,
  L5 (2004).

\bibitem[\protect\citeauthoryear{Watson et al.}{2004}]{whl+04}
  D. Watson, J. Hjorth, A. Levan, et al., Astrophys. J., 605, L101 (2004). 

\bibitem[\protect\citeauthoryear{Watson et al.}{2006}]{wvw+06}
  D. Watson, S.A. Vaughan, R. Willingale, et al., Astrophys. J. 636,
  967 (2006). 

\bibitem[\protect\citeauthoryear{Winkler et al.}{2003}]{wcd+03}
  C. Winkler, T.J.-L. Courvoisier, G. Di Cocco, et al.,
  Astron. Astrophys., 411, L1 (2003).

\end{thebibliography}
\end{document}